\documentstyle[epsfig,natbib2,natbibmnfix]{mn}
\newcommand{\afof}{{\sc AFOF}}

\newcommand{\fof}{{\sc FOF}}
\newcommand{\hop}{{\sc HOP}}

\newcommand{\mzero}{$M_{0}$}

\newcommand{\rt}{$r_{t}$}
\newcommand{\skid}{{\sc SKID}}
\newcommand{\sv}{$\sigma_{v}$}
\newcommand{\zt}{$\zeta_{t}\,$}
\newcommand{\be}{\begin{equation}}
\newcommand{\ee}{\end{equation}}
\newcommand{\bea}{\begin{eqnarray}}
\newcommand{\eea}{\end{eqnarray}}
\newcommand{\bc}{\begin{center}}
\newcommand{\ec}{\end{center}}
\newcommand{\bvbt}{\begin{verbatim}}
\newcommand{\evbt}{\end{center}}
\newcommand{\hkpc}{$h^{-1}$kpc}

\newcommand{\svm}{$\sigma_{v}-M_{0}$}
\newcommand{\simul}{16ML\_1}
\newcommand{\pld}{${\rm P}(\lambda )$}
\def\spose#1{\hbox to 0pt{#1\hss}}
\newcommand{\lta}{\mathrel{\spose{\lower 3pt\hbox{$\mathchar"218$}}
     \raise 2.0pt\hbox{$\mathchar"13C$}}}
\newcommand{\gta}{\mathrel{\spose{\lower 3pt\hbox{$\mathchar"218$}}
     \raise 2.0pt\hbox{$\mathchar"13E$}}}

\def\H0{$H_0 = 100~h~$km\,s$^{-1}$\,Mpc$^{-1}$}
\def\msun{$h^{-1}{\rm M}_{\odot}$}
\newcommand{\hmpc}{$h^{-1}$Mpc}

\bibpunct[,]{(}{)}{;}{a}{}{,}

\newif\ifAMStwofonts

\ifoldfss
  \ifCUPmtlplainloaded \else
    \NewTextAlphabet{textbfit} {cmbxti10} {}
    \NewTextAlphabet{textbfss} {cmssbx10} {}
    \NewMathAlphabet{mathbfit} {cmbxti10} {} 
    \NewMathAlphabet{mathbfss} {cmssbx10} {} 
  \fi
  \ifAMStwofonts
    \ifCUPmtlplainloaded \else
      \NewSymbolFont{upmath} {eurm10}
      \NewSymbolFont{AMSa} {msam10}
      \NewMathSymbol{\upi}     {0}{upmath}{19}
      \NewMathSymbol{\umu}     {0}{upmath}{16}
      \NewMathSymbol{\upartial}{0}{upmath}{40}
      \NewMathSymbol{\leqslant}{3}{AMSa}{36}
      \NewMathSymbol{\geqslant}{3}{AMSa}{3E}

      \let\leq=\leqslant 
      \let\geq=\geqslant \let\ge=\geqslant
    \fi
  \fi
\fi 

\ifnfssone
  \newmathalphabet{\mathit}
  \addtoversion{normal}{\mathit}{cmr}{m}{it}
  \addtoversion{bold}{\mathit}{cmr}{bx}{it}
  \newmathalphabet{\mathbfit} 
  \addtoversion{normal}{\mathbfit}{cmr}{bx}{it}
  \addtoversion{bold}{\mathbfit}{cmr}{bx}{it}
  \newmathalphabet{\mathbfss} 
  \addtoversion{normal}{\mathbfss}{cmss}{bx}{n}
  \addtoversion{bold}{\mathbfss}{cmss}{bx}{n}
  \ifAMStwofonts
    \ifCUPmtlplainloaded \else
      \UseAMStwoboldmath
      \makeatletter
      \new@mathgroup\upmath@group
      \define@mathgroup\mv@normal\upmath@group{eur}{m}{n}
      \define@mathgroup\mv@bold\upmath@group{eur}{b}{n}
      \edef\UPM{\hexnumber\upmath@group}
      \new@mathgroup\amsa@group
      \define@mathgroup\mv@normal\amsa@group{msa}{m}{n}
      \define@mathgroup\mv@bold\amsa@group{msa}{m}{n}
      \edef\AMSa{\hexnumber\amsa@group}
      \makeatother
      \mathchardef\upi="0\UPM19
      \mathchardef\umu="0\UPM16
      \mathchardef\upartial="0\UPM40
      \mathchardef\leqslant="3\AMSa36
      \mathchardef\geqslant="3\AMSa3E

      \let\leq=\leqslant 
      \let\geq=\geqslant \let\ge=\geqslant
    \fi
  \fi
\fi 

\ifnfsstwo
  \DeclareMathAlphabet{\mathbfit}{OT1}{cmr}{bx}{it}
  \SetMathAlphabet\mathbfit{bold}{OT1}{cmr}{bx}{it}
  \DeclareMathAlphabet{\mathbfss}{OT1}{cmss}{bx}{n}
  \SetMathAlphabet\mathbfss{bold}{OT1}{cmss}{bx}{n}
  \ifAMStwofonts
    \ifCUPmtlplainloaded \else
      \DeclareSymbolFont{UPM}{U}{eur}{m}{n}
      \SetSymbolFont{UPM}{bold}{U}{eur}{b}{n}
      \DeclareSymbolFont{AMSa}{U}{msa}{m}{n}
      \DeclareMathSymbol{\upi}{0}{UPM}{"19}
      \DeclareMathSymbol{\umu}{0}{UPM}{"16}
      \DeclareMathSymbol{\upartial}{0}{UPM}{"40}
      \DeclareMathSymbol{\leqslant}{3}{AMSa}{"36}
      \DeclareMathSymbol{\geqslant}{3}{AMSa}{"3E}

      \let\leq=\leqslant 
      \let\geq=\geqslant \let\ge=\geqslant
    \fi
  \fi
\fi 

\ifCUPmtlplainloaded \else
  \ifAMStwofonts \else 
    \def\upi{\pi}
    \def\umu{\mu}
    \def\upartial{\partial}
  \fi
\fi

\title{Properties of galaxy halos in Clusters and Voids}
\author[V. Antonuccio-Delogu et al.]
  {V.~Antonuccio-Delogu,$^1$\thanks{Affiliated to: Theoretical Astrophysics Center, Copenhagen, Denmark .} 
  U.~Becciani, $^1$  E.~van Kampen,$^2$ A.~Pagliaro,$^3$ A. B. Romeo,$^4$ 
\newauthor
S.~Colafrancesco,$^5$ A. German\'{a}, $^1$ M. Gambera $^6$\\
  $^1$Osservatorio Astrofisico di Catania, Citt\'{a} Universitaria, Via Santa Sofia 78,
95123 Catania, Italy\\
  $^2$Institute of Astronomy, University of Edinburgh, Edinburgh EH9 3HJ, 
United Kingdom\\
  $^3$SRON, 3584 CA Utrecht, The Netherlands\\
$^4$Department of Astronomy and Astrophysics, Centre for Astrophysics and
Space Science,  Chalmers University of Technology\\
SE-43992 Onsala, Sweden \\
$^5$Osservatorio Astronomico di Roma, Monteporzio Catone, Italy\\
$^6$Liceo Scientifico Statale ``E. Majorana'', Scordia, Italy
}
\date{Accepted ???
      Received ???;
      in original form ???}

\pagerange{\pageref{firstpage}--\pageref{lastpage}}
\pubyear{2000}

\begin{document}

\maketitle

\label{firstpage}

\begin{abstract}
We use the results of a high resolution N-body simulation to investigate 
the r\^{o}le of the environment on the formation and 
evolution of galaxy-sized halos. Starting from a set of constrained
initial conditions, we have produced a final configuration hosting a
double cluster in one octant and a large void extending over two
octants of the simulation box. In this paper we concentrate
on {\em gravitationally bound} galaxy-sized halos extracted from these two
regions and from a third region hosting a single, relaxed cluster without
substructure.  Exploiting the high mass resolution of our simulation
($m_{body} = 2.1\times  10^{9} h^{-1} M_{\odot}$), we construct halo samples
probing more than 2 decades in mass, starting from a rather small mass
threshold:  $5\times 10^{10} h^{-1} M_{\odot}\leq M $.  We present results for two
statistics: the relationship between 1-D velocity dispersion \sv\, and mass
\mzero\, and the probability distribution  of the spin parameter $P(\lambda )$,
and for three different group finders. The \svm\, relationship is well
reproduced by the  Truncated Isothermal Sphere (TIS) model introduced by
\citet{1999MNRAS.307..203S}, although the slope is different from the original
prediction. A series of \svm\, relationships for different values of the anisotropy
parameter $\beta$, obtained using the theoretical predictions by \citet{2001MNRAS.321..155L}
for \citet{1996ApJ...462..563N, 1997ApJ...490..493N} density profiles are found to be only
marginally consistent with the data.
Using some properties of the equilibrium TIS models, we construct
subsamples of {\em fiducial} equilibrium TIS halos from each of the three
subregions, and we study their properties. For these halos, we do find an
environmental dependence of their properties, in particular of the spin
parameter distribution $P(\lambda )$. We study in more detail the TIS  model,
and we find new relationships between the truncation radius and other
structural parameters. No gravitationally bound halo is found having a radius
larger than the critical value for gravithermal instability for TIS halos
(\rt\, $\ge 34.2 r_{0}$, where $r_{0}$ is the core radius of the TIS solution).
We do however find a dependence of this relationship on the environment, like
for the $P(\lambda )$ statistics. These facts hint at a possible r\^{o}le of
tidal fields at determining the statistical properties of halos.

\end{abstract}

\section{Introduction} \label{intr}
One of the distinguishing features of any scenario for the formation of the Large
Scale Structure of the Universe within the Cold Dark Matter (CDM) cosmological
model is represented by the {\em hierarchical clustering} paradigm for the 
assembly of gravitationally bound structures \citep{1996grdy.conf..121W,
1997evun.work..227W}. In its
simplest form, the idea of hierarchical clustering implies the fact that the growth of 
halos proceeds by accretion of smaller units from the surrounding environment,
either by infall \citep{1972ApJ...176....1G} or by a series of of ``merging`` events 
\citep{1978MNRAS.183..341W}, whereby the subunits are accreted in a discontinous way, or
(more likely) by a combination of the two. In the first case the typical halo profiles evolve
adiabatically, while in the merging scenario each merging ``event'' will induce some transients in
the characteristic properties which in turn will induce some evolution in the
typical profiles, after the subunits have been accreted and destroyed. In
either case, one expects that relaxation processes should drive  the evolution
towards a quasi-equilibrium state on a dynamical timescale, a state
characterized by  relationships among global quantities related to the halo,
like its mass, density, velocity  dispersion \sv , and possibly others.
Recently a considerable attention has been devoted to the study of one of 
these relationships: the radial dependence of the (spherically averaged)
density, also known as the density profile. Unfortunately the density profile
is a very difficult tool to use when trying to characterize the statistical
properties of halo populations, because the predictions of different models of
halo formation differ only in the behaviour in the central parts, where the
statistics is typically poor. Less attention has been paid
to another global quantity, namely the velocity dispersion and to its
relationship with other global quantities, like the mass. The velocity
dispersion enters the second-order Jeans equation, while the density profile is
described by the zeroth-order Jeans equation \citep{1987gady.book.....B}. For
this reason it contains different physical information than the density
profile. Recently \citet{1998ApJ...495...80B} have looked at the \svm\,
relationship for clusters, and they find a good agreement with the standard,
singular isothermal sphere model as far as the slope of the relationship is
concerned. Also \citet{1999A&A...341....1K} looked at this relationship, using
a different code. Halo equilibrum models make predictions about the \svm\,
relationship, but these are difficult to compare with observations, because
some of the involved quantities (e.g. the velocity dispersion itself) are not
directly deducible from observations. They can however be studied with N-body
simulations, and one of the purposes of this paper will be to show that the
\svm\, relationship can be used to discriminate among different halo
equilibrium models.\\ A second problem we will study  concerns the dependence
of halo properties on the environment within which they form. In both the
hierarchical clustering scenarios mentioned above  one could imagine that the
properties of the halos do depend on the environment. For instance, the
dynamics of the infall process could be affected by the average overdensity of
the environment within which the halo grows \citep{1977ApJ...218..592G}, or by
its shear \citep{1999astro-ph...9912347, 1999GReGr..31..461T}. Also typical
quantities  related to the merging, like the frequency of merging events, could
intuitively be affected by the average density of the environment, at least for
galaxy-sized halos forming within clusters. High resolution N-body simulations
\citep[e.g.][]{1999ApJ...524L..19M} show that most of the galaxies lying in the
central (i.e. virialised) parts of clusters do not easily reach a velocity
larger than the escape velocity: so, they are bound to the cluster for the
largest part of their evolutionary history, and consequently form in an
overdense environment. It would then be interesting to try to understand
whether there are systematic differences between halos forming in clusters and
in voids.\\  Some of these issues have been recently discussed in the
literature. \citet{1999MNRAS.302..111L} have analysed the dependence of various
statistical properties, including the spin probability distribution, on the
environment, and found no evidence for any dependence apart for the extent of
the mass spectrum. They divide their halos in groups according to the
overdensity of the environment within which they are found, and show that the
scatter diagrams between different quantities are indistinguishable among the
different groups. In the present study, we follow a different strategy. We
study a simulation obtained from a constrained set of initial conditions, in
order to get a few clusters (and, in particular, a {\em double cluster}) and a
large void within the same simulation box. We then extract our halos from three
spatially disjoint regions: one containing a double cluster, a second one contaning
a single cluster a third one containing the void. This
is in some sense  complementary to the procedure which
\citeauthor{1999MNRAS.302..111L} seem to  have followed, because our halos are
grouped according to the spatial distribution, rather than according to the
overdensity, so they are grouped according to the {\em environment} within
which they form.\\ Very recently \citet{2000astro-ph...0006342} presented a
study of the spin probability distribution for 6 different cosmological models
and environments. He finds a difference between the distributions of halos
resulting from recent mergers and halos which did not experience mergers,
almost  independent of the environment within which they form. This could have
significant consequences for the construction of merger histories, and,
ultimately, for the semi-analytical modelling of galaxies. Similar results have been 
recently obtained by \citet{2001astro-ph...0105349}.\\ 
The plan of the paper is as follows. In section ~\ref{sims} we describe the numerical setup of
the simulations and the algorithm adopted to identify halos. In section
~\ref{hem} we describe the halo equilibrium models with which we compare the
results of our simulation, and in section ~\ref{sp} we show the results of this
comparison and discuss their physical interpretation. Finally, in the 
conclusions we summarize our results and suggest some directions for future
studies.

In the following we will always assume a $\Omega = 1$ Standard Cold Dark
Matter model, with a Hubble constant $H_{0}= 100 h \, {\rm Km\, s^{-1}\,
Mpc^{-1}}$, and $h=0.5$. All lengths, unless explicitly stated, are assumed to be
comoving.

\section[]{Simulations} \label{sims}
The simulation from which the data have been extracted has been performed using
FLY \citep{1997CoPhC.106..105B,1998adass...7....7B} a parallel, collisionless treecode
optimised for Shared Memory and/or clustered computing systems.
FLY deals with periodic boundary conditions using a standard Ewald summation
technique \citep{1991ApJS...75..231H}. The algorithm adopted is the octal-tree algorithm of
Barnes \& Hut \citep{1986Natur.324..446B}, with some modifications
 (``grouping'' of cells belonging to the lists of nearby particles,
\citealt{1987CoPhC.87..161B}) during the phase of tree walking. These changes have a
negligible impact on the overall numerical accuracy, as shown elsewhere
\citep{2000JCoPh}, but they have a strong positive impact on parallel
peformance and scaling.\\  
We have performed two simulations starting from the
same initial conditions. In both cases the underlying cosmological model is a
Standard Cold Dark Matter (SCDM), with $\Omega_{0}=1, \sigma_{8}=0.9$. The main
reason for this choice lies in the fact that the specific prediction for the
\svm\, statistics we consider in the next
sections were done for this particular cosmological model. We plan to extend
our work to other cosmological models in future work.\\ 
Each simulation used
$256^{3}$ particles, and the box size was 50\hmpc, so that the mass of each
particle is $m_{part}=2.07\times 10^{9}$ \msun. The simulations were designed
to study the evolution of a Coma-like cluster, and for this purpose constrained
initial conditions were prepared, changing only the softening length, which was
fixed to $\epsilon=10$ \hkpc\, and $5$  \hkpc , respectively. As far as the
results presented in this paper are concerned, there are no differences among
these two simulations, so for the rest of this paper we will concentrate only
on the simulation with the largest softening length, which we designate in the
following as \simul.

\begin{figure}
\label{fig_box}
\psfig{figure=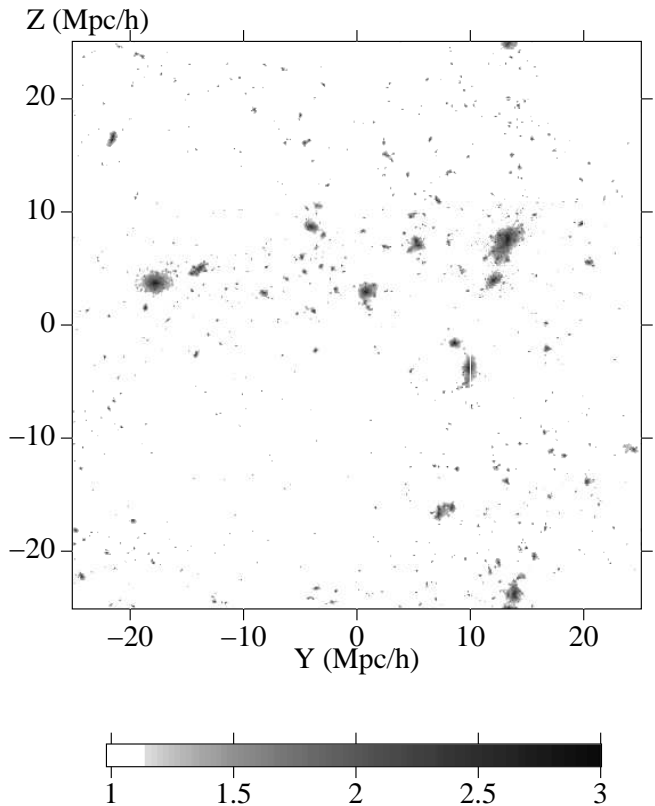,width=9.0cm}
\caption{Snapshot of the simulation box at the end of the simulation. The scale of
grey corresponds to density in logarithmic units. The large void is clearly seen in
projection extending over the lower left octant.} 
\end{figure}
Constrained initial conditions were prepared using the implementation of the constrained random
field algorithm of \citet{1991ApJ...380L...5H} by \citet{1996MNRAS.281...84V}. We took the same
initial conditions adopted in one of the simulations from the catalogue of
\citet{1997MNRAS.289..327V}.  More specifically, we constrain the initial conditions to have a
peak at the centre of the simulation box, with height $3.04\sigma$,
Gaussian smoothed at a scale of $6 h^{-1}$ Mpc, and a $-2\sigma$
void centered at $(15,0,-10)$.
The final configuration is shown in Figure~1. In order to study the environmental dependence of the
properties of galaxy-sized halo populations, we selected three regions within the
computational box, which we call DOUBLE, SINGLE and VOID. All the three are
cubical  with centers and sizes as specified in Table~\ref{psim}. The DOUBLE
region hosts a double cluster, with two large parts in the act of merging by the end
of the simulation (see Fig.~2). The SINGLE region hosts a more relaxed cluster
without any apparent substructure. Finally, we have included in the analysis a
significantly underdense region, the VOID, which is more extended than the
former two, so as to contain enough halos to allow a reasonable statistics.

\begin{figure}
\label{fig_double}
\psfig{figure=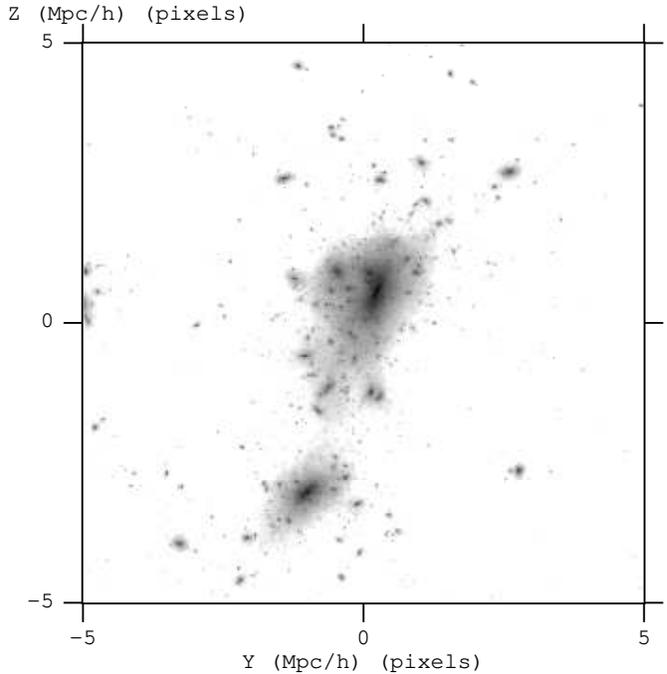,width=9.0cm}
\caption{The region of the DOUBLE cluster.} 
\end{figure}

\begin{table}
 \caption{Properties of the analyzed regions.}
 \label{psim}
 \begin{tabular}{@{}lccccc}
  Region & $x$ & $y$ & $z$ 
       & $L$  & Number of halos \\
 & & & & & \\
  DOUBLE & 15 & 12.5
       & 7.5 & 10 & 827 \\
  SINGLE & 12 & -18
       & 4 & 10 & 786 \\
   VOID & -10 & 
       -10. & -10.  & 20 & 609 \\
  
 \end{tabular}

 \medskip
 All lengths are in \hmpc. From left to right, columns are as follows: label
of the region, x,y,and z coordinate of its center, size of the region, total number of
halos found. 
\end{table}

\subsection[]{Finding halos} \label{fh}
Various methods have been devised to extract halos from the outputs of N-body
simulations. Some of these methods make use only of particle positions, like
the standard Friends-Of-Friends (herefater FOF) and the various versions of the
Adaptive FOF, while others take into account also particle velocities
\citep[e.g. $SKID$,][]{1997NewA....2...91G} and/or environmental
properties like local densities \citep[like \hop,][]{1998ApJ...498..137E}. Most
of the results we will present later have been obtained using 
$SKID$ because it selects {\em gravitationally
bound} groups of particles. In short, $SKID$ first builds catalogues of groups
using a standard FOF algorithm, selecting only
particles lying within regions whose density is larger than a critical value
$\delta_{crit}$.  It then computes the escape velocity of each particle and discards 
those particles having a rms velocity larger than the escape
velocity. This ``pruning'' procedure should then leave only those particles
which are actually bound to the group, discarding those ``background''
particles which find themselves by accident at  a given time within it. The
initial linking length of the FOF phase determines the approximate size of the
groups we are considering. We assumed a linking length of $100 h^{-1} {\rm
kpc}$, corresponding to the typical size of a galaxy-sized object at the
present epoch. The softening length for the calculation of the gravitational
potential was assumed to be the same as in the simulation, and the critical
density $\delta_{crit}$ was set equal to $178/(1+z)$, the value for nonlinear
collapse in the Gunn \& Gott collapse model, so that only particles from
genuinely nonlinearly collapsed shells should be included.\\ 
Following a suggestion of the anonymous referee, we have also adopted two more
halo finders to check the robustness of the results: an Adaptive
FOF method devised by \citet{1997MNRAS.289..327V} and a modified version of
\skid\, which should avoid the problems posed by the original version. 
This particular AFOF halo finder selects only those haloes that 
are virialized, by specifically testing for virialization.
Concerning the second method, we have modified \skid\, only in that part which
engenders the input group list which is subsequently ``pruned'' of the
non-gravitationally bound particles: in place of a standard FOF (as in the
original version of \skid ) we have adopted \hop \citep{1998ApJ...498..137E} as
input group finder. As we will see later, some of the relationships we find do
depend on the group finder adopted, but those relationships holding for the
equilibrium TIS halos are not affected by this. 
All the results we present in the following are for the final redshift of the
simulation, $z=0.0047$, unless otherwise stated.

 \section[]{Halo equilibrium models} \label{hem}

The internal properties of halos formed by gravitational collapse can
be described by looking at correlations among different physical
quantities. The density profile has
often been used to study the properties of relaxed, virialised halos,
particularly since the findings by \citet*{1996ApJ...462..563N} that this
profile has a {\em universal} character when expressed in dimensionless units. 
However, the density profile can be reliably determined only for halos having
enough particles in each shell to minimise the statistical fluctuations.
For instance, \citet{1997ApJ...490..493N} considered only 8 halos
extracted from a low resolution simulation and re-simulated with a higher mass
resolution.\\
Typical N-body simulations on cluster scales tend to produce a large amount of 
halos, whose density profile can not be reliably determined, because each of them
contains on average less than $10^{5}$ particles. For this
reason we have chosen to study relationships involving {\em global} halo
properties. This choice is not free from potential problems: systematic biases
can be introduced by the particular group finder adopted. Consider for instance
\skid, which works by stripping out gravitationally unbound particles from
halos builded using FOF: the group catalogues so produced tend
to be more biased towards less massive halos than catalogues produced using
FOF. We have then decided to adopt three different group finders, in order to
be able to understand the role of these systematic factors. We
have also considered two different statistics to characterize the properties of
dark matter halos, and particularly their equilibrium properties at the end of
the simulation: the internal 1-dimensional velocity dispersion - mass
relationship (\svm\,) and the spin probability distribution ${\rm P}(\lambda
)$. Theoretical predictions concerning both of them are available in the
literature. In particular, we will compare the results from our N-body
simulations with four models: the Standard Uniform Isothermal Sphere (SUS)
\citep[see e.g. ][chap.~8 for a detailed treatment]{1993sfu..book.....P}, the 
Truncated Isothermal Sphere models (TIS) recently introduced by 
\citet{1999MNRAS.307..203S}, the ``peak-patch'' (PP) Montecarlo models by
\citet{1996ApJS..103....1B}, and some models derived from the 
\citet{1996ApJ...462..563N, 1997ApJ...490..493N} (hereafter NFW) density profiles. 
The first two models predict a \svm\, relationship given by: 
\be 
\label{eq_sv} \sigma_{v} = c_{f} \,{\rm
M}_{12}^{1/3} \left( 1+z_{coll}\right)^{1/2}h^{1/3}\, {\rm Km\, s^{-1}}
\label{hem:eq_1} \ee 
where the subscript $f\equiv SUS \, {\rm or}\,  TIS$,
while for the PP model the relation is given in \citet{1996ApJS..103...41B}:
\be 
\sigma_{v} = c_{PP} \,{\rm M}_{12}^{0.29} \left(
1+z_{coll}\right)^{1/2}h^{1/3}\, {\rm Km\, s^{-1}} \label{hem:eq_2} 
\ee 
In the equations above ${\rm M}_{12}$ is the mass in units of $10^{12}\, {\rm
M}_{\odot}$ and $z_{coll}$ is the collapse redshift. The coefficients for these
cases are given by \be c_{SUS, TIS, PP}=(71.286, 104.69, 117.60)
\label{hem:eq_3} \ee respectively. We restrict our attention to these four
models because the physical ingredients which enter in their formulation are
very different, and encompass a sufficiently wide range among all the possible
nonlinear collapse and virialization mechanisms. This wide choice reflects our 
generally poor level of understanding of the nonlinear physics of gravitational
collapse, of its  dependence on the local environment and on other properties
like the merging history of the substructures.\\ The SUS model is based on the
spherical nonlinear collapse model  \citep{1972ApJ...176....1G}. In this model 
the collapse towards a singularity of a spherically simmetric shell of matter
in a cosmological background is halted when its radius reaches a value of half
the maximum expansion radius. The velocity dispersion is then fixed by imposing
the condition of energy conservation, which must hold in the case of
collisionless dark matter as that envisaged here. The TIS model also considers
the highly idealized case of a spherically symmetric configuration, but assumes
that the final, relaxed system is described by an isothermal, isotropic 
distribution function and that the density profile is truncated at a
finite radius. \citet{1999MNRAS.307..203S} have shown that this configuration could
arise from a top-hat collapse of an isolated spherical density
perturbation if, as shown by \citet{1985ApJS...58...39B}, the
dimensionless region of shell crossing almost coincides with the region
bounded by the outer shock in a ideal gas accretion collapse with the
same mass (in a $\Omega=1$ CDM model). The truncation radius is
then assumed to coincide with the region of shell crossing, and this
allows them to specify completely the model.\\
The peak-patch models introduced by \citet{1996ApJS..103....1B}
are more general than the SUS model, in that they include a more realistic collapse model where
deviations from spherical symmetry are taken into account. The density
perturbation is approximated as an axisymmetric homogeneous spheroid. Coupling
between the deformation tensor and the external and internal
torques are consistently taken into account up to a few first
orders, and Montecarlo realizations are used to build up
catalogues of halos. These have been compared with the N-body
simulations of \citet{1991ApJ...368L..23C} in order to properly normalize the statistics. Note that
eq.~\ref{hem:eq_2} is a best-fit relationship and holds for a range of halo mass
($2.5\times 10^{14}\leq M_{h}\leq 5\times10^{15} {\rm M}_{\odot}$) much larger 
than that considered here. Nonetheless, we  include it into our 
comparison because the physical model it is based on is significantly different 
from the other models we consider.\\
Finally, we have considered models for the \svm\, statistics consistent with the 
NFW density profile, which were recently derived by \citet{2001MNRAS.321..155L}
solving the second order Jeans equation:
\be
    \frac{1}{\rho} \frac{\rm d}{{\rm d} r} (\rho \sigma_{\rm r}^2) +
    2 \beta \frac{\sigma_{\rm r}^2}{r} = -\frac{{\rm d} \Phi}{{\rm d} r} \ , \label{eq_nfw1}
\ee
where $\beta=1-\sigma_\theta^2(r)/\sigma_{\rm r}^2(r)$ quantifies the anisotropy 
of the velocity dispersion. For a NFW density profile we have:
\be
\rho_{NFW}(r)\equiv \rho_{NFW}(s) = \frac{\rho_{c}^{0}c^{2}g(c)}{3}\cdot\frac{1}{s\left( 1+cs\right)^{2}}
\label{eq_nfw2}
\ee
\be
\Phi_{NFW}(s) = -\frac{GM_{VIR}}{r_{VIR}}\cdot g(c)\frac{\ln\left( 1+cs\right)}{s} \,
\label{eq_nfw3}
\ee
where we have defined: $r_{VIR}, M_{VIR}$ as the virialisation radius and mass, respectively, 
$s=r/r_{VIR}$, $c=c(M,z)$ is the concentration parameter and: $g(c)=1/(\ln (1+c) - c/(1+c))$.
Eq.~\ref{eq_nfw1} can be solved by quadrature, and the solution finite in the limit $r\rightarrow \infty$
is:
\begin{eqnarray}
    \frac{\sigma_{\rm r}^2}{V_v^2} (s, \beta={\rm const})
    &=& g(c) (1+c s)^2 s^{1-2 \beta}
    \nonumber \\
    &\hspace{-2cm} \times & \hspace{-1.4cm} \int_{s}^\infty
    \left[ \frac{s^{2 \beta - 3} \ln (1+c
    s)}{(1+c s)^2} - \frac{c s^{2 \beta-2}}{(1+c s)^3} \right] {\rm d} s .
    \label{eq_nfw4}
\end{eqnarray}
\noindent
Note that we have always chosen as critical treshold for our group finders the virialisation overdensity
($\delta \geq \delta_{crit}=178$), so the quantities $r_{VIR}, M_{VIR}$ are the actual radius and mass
found by the group finders for each halo.\\
In order to make use of eq.~\ref{eq_nfw4} we have yet to specify the dependence of the concentration 
parameter on the mass at the final redshift: $c=c(M,z=0)$. We adopt the relationship provided by 
\citet{2001MNRAS.321..559B}, by running their code CVIR for the relevant cosmological model. In the mass range 
we are interested to ($5\times 10^{10} \leq M/M_{\odot} \leq 5\times 10^{13}$) we find a power law fit:
$c(M,z=0) = 472.063\times M^{-0.127 \pm 0.01}$. Finally, in order to make a proper comparison with the
quantity computed by the group finder, we evaluate the mass averaged velocity dispersion:
\be
\sigma_{v}^{2} = \frac{4\pi \int_{0}^{1}\sigma_{v,NFW}^{2}(s)\rho(s)s^{2}ds}{M(1)}   \label{eq_nfw5}
\ee
where we have defined:
\[
\sigma_{v,NFW}^{2} = \sigma_{r}^{2} + \sigma_{\theta}^{2} = (2-\beta) \sigma_{r}^{2}
\]
and:
\[
M(1)= 4\pi \int_{0}^{1}\rho(s)s^{2}ds
\]

\section[]{Statistical Properties} \label{sp}

\subsection[]{The \svm\, relation} \label{sv_m}
In Figure~3 we show plots of the final
\svm\, relationship for galaxy-sized halos in the 
DOUBLE, SINGLE and VOID regions, respectively, obtained using \skid . The most striking  
difference is probably the different character
of halos in the VOID  region when compared with the clustered regions. Halos in
VOID  have a much smaller dispersion around the mean than halos in the 
clustered regions, and the distribution is almost symmetric with
respect to the best fitting approximation.
Halos in the DOUBLE region have a larger
dispersion and they do show an asymmetry in the distribution 
around the best fit solution, i.e. an excess of low mass halos at low
$\sigma_{v}$. The latter point is useful to understand the potential
systematic effects introduced by a particular group finder. In Figure~4 we show
the \svm\, relationship for the DOUBLE region obtained by using the two
other group finders mentioned above. As is evident, the excess of low-mass halos
is only an artifact introduced by \skid, which makes use of a \fof \,
algorithm to build up an input list of groups. The main results of the next
sections, however, do not depend on the particular group finder adopted, 
because we will select subsamples of halos which can be regarded as {\em
equilibrium TIS} halos, and for them the slope of the \svm\, relationship is
independent of the particular group finder initially adopted.\\  
A more important difference is evident from a comparison between clustered and
void regions. Halos in VOID have a larger mass extent \citep[a property
already noted by ][]{1999MNRAS.302..111L} and also the slope of the \svm\,
relation seems to be larger than for the other two cases.\\
In Figure 5 we show a comparison with some theoretical predictions for NFW
density profiles. In order to apply eqs.~\ref{eq_nfw3},~\ref{eq_nfw4} we have
still to specify a relationship between the virial radius and the corresponding
mass, which enters into eq.~\ref{eq_nfw3}. We do this by fitting a power law
relationship to the data obtained from the simulations (Figure 7), which shows
that the slope depends signficantly on the environment. As is evident from
Figure 5, none of the models fits adequately the data. Note that there is only
a slight difference bewteen isotropic ($\beta = 0$) and anisotropic ($\beta =
0.5$) models. Only if we allow for an unrealistically low value of the slope of
the $r_{VIR}-M_{VIR}$ relationship we get a limited agreement for the DOUBLE
cluster, but not for the other two regions.\\
The slopes of the \svm\, relationship for different regions and using
different group finders seem to be consistent with each other, within the
errors (Table \ref{svtab}). None of the theoretical models we are
considering, however, seems to offer a good fit for all the cases. The TIS
model seems to give a good fit for the DOUBLE region and for the \skid\, group
finder, but when we use the modifed \skid\, group finder, which
produces a sample over a more extended mass interval, we see that the original
TIS model does not offer a good fit (Figure~4). 

Note that the rms uncertainty of \sv\, in Figures~3-7 is less than 30
Km/sec, a value much lower than that found by \citet{1999A&A...341....1K} in
their simulations (see their Fig. 3). We believe that this is a consequence of
the larger mass and force resolution of our simulation, and also of the use
of a larger dynamic range than adopted by previous authors.
 
Generally speaking, a power-law seems to offer a good fit for all the
three regions (although with
different ranges for the three regions), but in order to determine the slope
one must probably go a step further in modelling the physical state of these
halos. In the next section we will explore in more detail the properties of TIS
halos and we will focus our attention on their statistical properties.

\begin{table} 
\caption{Least-square best fit parameters for \sv\, relationship. }  
\label{svtab}   \begin{tabular}{@{}lccccc}    Region & $\alpha $ &
$\Delta\alpha $ & $c_{0}$ & & Method  \\  & & & & & \\   DOUBLE & 0.38 & 0.08  
     & 74.13 &  & SKID \\  & 0.39 & 0.05 & 87.74 & & SKID with \hop input \\
 & 0.42 & 0.07 & 72.48 & & AFOF \\
 & 0.42 & 0.03 & 82.16 & & TIS selected halos\\
 & & & & & \\
  SINGLE & 0.35 & 0.04
       & 80.28 & &  SKID\\
 & 0.37 & 0.04 & 82.15 & & SKID with \hop input \\
 & 0.42 & 0.07 & 71.22 & & AFOF \\
 & & & & & \\
  VOID & 0.39 & 0.04
       & 86.81 &  & SKID \\
 & 0.40 & 0.04 & 89.43 & & SKID with \hop input \\
 & 0.45 & 0.05 & 63.12 & & AFOF \\
 & 0.38 & 0.02 & 88.60 & & TIS selected halos \\
  
 \end{tabular}

 \medskip
$\alpha, c_{0}$ are the fitting parameters of a
power law fit of the form: $\sigma_{v}=c_{0}M_{12}^{\alpha}$, $\Delta\alpha$
is the rms error associated with $\alpha$.
\end{table}
\begin{figure}
\label{fig_sm_2d}
\vspace*{-3cm}
\psfig{figure=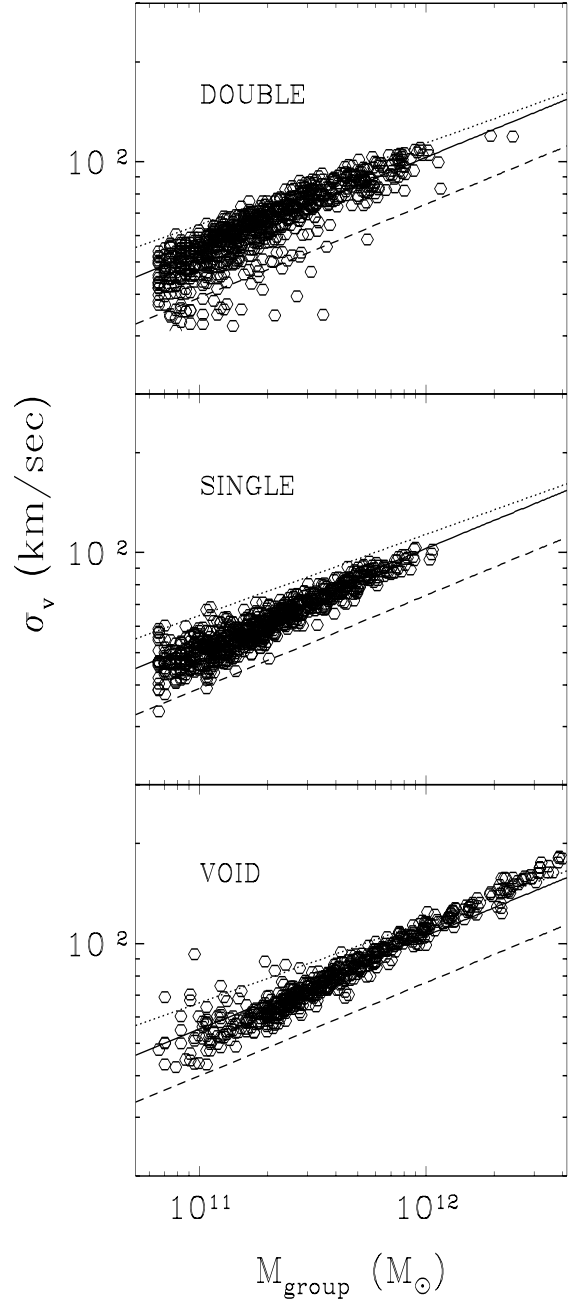,height=21cm,width=8.0cm}
\caption{1-D velocity dispersion versus mass for halos extracted from the three
regions. The three fitting curves corrsond to the three cases considered in
the text: Truncated Isothermal Sphere (continuous line), Bond \& Myers 1996
(dotted line), Standard Uniform Sphere (dashed line). Note the larger mass
extent for halos in the VOID. In all the cases, the slope is larger than
predicted by models, although within clusters the statistical uncertainty is
large.}    
\end{figure}

\begin{figure}
\label{fig_sm_hop}
\psfig{figure=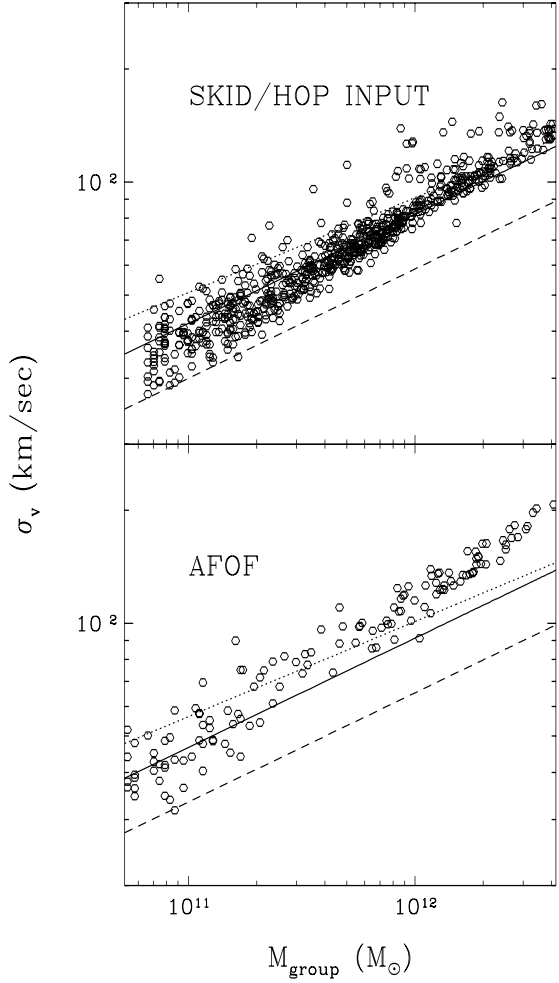,height=14cm,width=8.0cm}
\caption{1-D velocity dispersion versus mass for the DOUBLE cluster region,
halos selected using the AFOF and modified \skid\, with \hop input group
finders. The same parameters adopted for \skid\, in Figure~3 are adopted.
The best fit power law for the plot in the upper figure has a slope: 
$\alpha=0.039\pm 0.05$. }  \end{figure}

\begin{figure} \label{fig_nfw} 
\psfig{figure=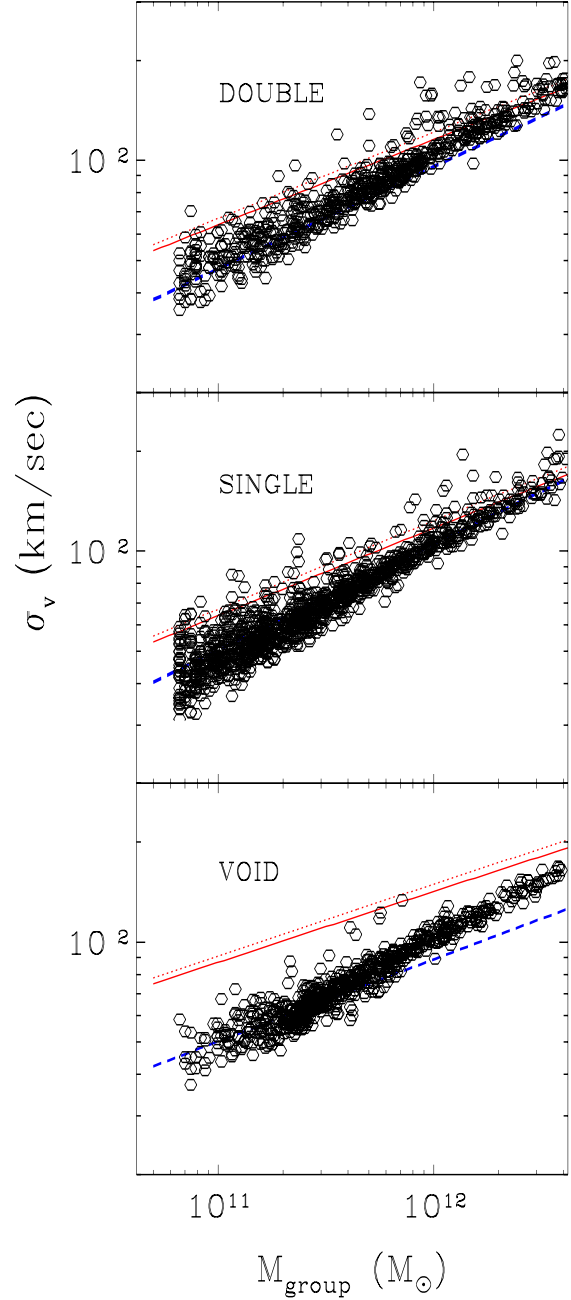,height=21cm,width=8.0cm}
\caption{The \svm\, relationship for three models based on the NFW density
profile. Continous line: $\beta=0$, slope of the $r_{VIR}-M_{VIR}$
given by $\alpha=\alpha_{mean}$. Dotted line: $\beta=0.5, \alpha=\alpha_{mean}$.
Dashed line: $\beta=0, \alpha=\alpha_{mean} - \Delta\alpha$ (i.e. the 1$\sigma$
limit).
}  \end{figure}

\subsection[]{Comparison with the Truncated Isothermal Sphere model} \label{tis}
We will now consider the possibility of obtaining a reasonable fit of
the \svm\, relationship by modifying the minimum-energy TIS model. We will then 
present here some more features of this model.\\ 
Following \citet{1999MNRAS.307..203S}, the TIS solution is obtained by
imposing a finite truncation radius \rt\, on an isothermal, spherically
symmetric collisionless equilibrium configuration. Shapiro et al. define a
typical radius:

\be \label{r0def} r_{0}=\frac{\sigma_{v}}{\left( 4\pi
G\rho_{0}\right)^{1/2}} \label{eq1} 
\ee

where $\rho_{0}$ is the central density
(TIS models are non-singular). Combining the Poisson and the Jeans' equilibrium
equations, and making the hypothesis of isothermality for the distribution
function, they obtain an equation for the dimensionless density \citep[see][eq.
29]{1999MNRAS.307..203S}: 

\be \label{nondim_sph} 
\frac d{d\zeta}  \left(\zeta^2
\frac{d(\ln\tilde{\rho})}{d\zeta}\right) 	= -\tilde{\rho} \zeta^2. 
\ee

where we have used the definitions:
\[
\label{rho0zetadef}
\tilde{\rho} = \displaystyle{\frac{\rho}{\rho_0}}, \qquad
\zeta = \displaystyle{\frac{r}{r_0}},
\]
Shapiro et al. have shown that nonsingular solutions of eq. \ref{nondim_sph} form a
one-parameter family depending only on $\zeta_t=r_{t}/r_{0}$. The total mass is then
given by:
\be
\label{mass0}
M_0=M(r_t)=\int^{r_t}_0 4\pi\rho(r) r^2d\,r=4\pi\rho_0r_0^3 \tilde{M}(\zeta_t)
\ee
where we have defined a dimensionless total mass:
\[
\tilde{M}(\zeta_t)
=\frac{1}{r_{0}} \int_{0}^{r_t} dr  {\rho \over \rho_{0}} \bigg({r \over
r_{0}}\bigg )^2
\]
We follow further \citet[][eq. 41]{1999MNRAS.307..203S} and
write the virial theorem for a collisionless truncated isothermal sphere:
\be
\label{vir_th}
0=2K+W+S_p,
\ee
where: $K=M_0 \langle v^{2}\rangle/2\equiv \left( 3/2\right) M_0\sigma_{v}^{2}$ and
$S_p$ is the surface pressure term. In the Appendix we show that, after some simple
algebra, starting from eq. \ref{vir_th} one obtains the following relation:
\be
\label{eq_glob}
\frac{GM_0}{r_{t}\sigma_{v}^{2}}=\frac{3\tilde{M}(\zeta_t)-\tilde{\rho}(\zeta_t)
}{\zeta_t\Psi(\zeta_t)}\tilde{M}(\zeta_t)\equiv\Phi(\zeta_t )
\ee
The function $\Psi(\zeta_t)$ is specified in the Appendix.\\
We have already seen that a power-law fit describes well the \svm\,
relationship. Then from eq. \ref{eq_glob} we deduce that also the truncation
radius \rt\, has a power -law dependence on the total mass: $r_{t}\propto c(\zeta_{t}
) M_{0}^{1-2\alpha}$. Using the values of $\alpha$ from Table \ref{svtab} we see
that the slope of this relationship should lie within the range 0.22-0.3, i.e. it should be quite
small. In Figure~7 we plot $r_{t}$ as a function of $M_{0}$. The
best-fit values we find for the slope are inconsistent with the predictions
from the \svm\, relationship. \\
The right hand side of equation~\ref{eq_glob} depends only on the
dimensionless truncation radius $\zeta_t$, which in the minimum-energy TIS
solution of \citet{1999MNRAS.307..203S} should be fixed to $\zeta_t=29.4$. The
function $\Phi(\zeta_t )$ has a singularity at $\zeta\approx 0.97$, where the
denominator goes to zero (Figure~8). However, we are interested in the region
$\zeta_t > 1$, where the truncation radius is at least comparable to the core
radius. As is clear from Fig.~10, the function has  minimum at $\zeta\approx
59.5$.

\begin{figure} \label{fig_zm} 
\psfig{figure=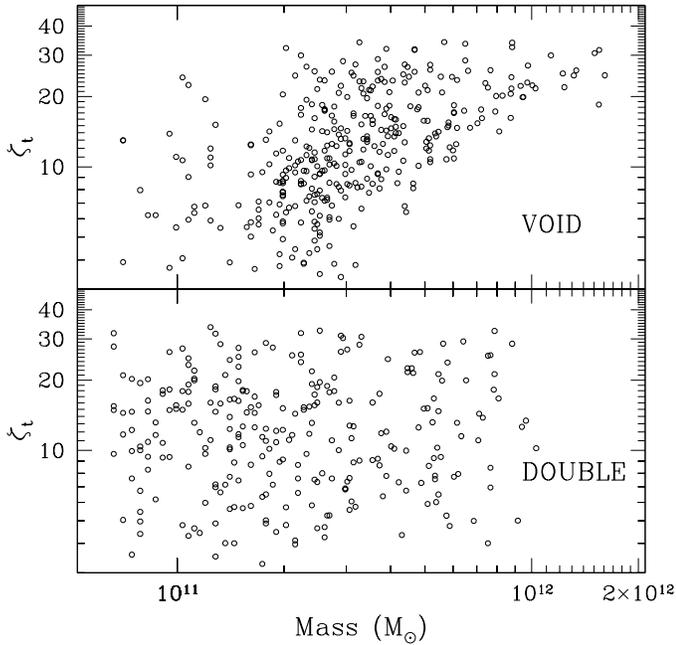,width=9.0cm}
\caption{The truncation parameter $\zeta_{t}$ as a function of group's mass.} 
\end{figure}

\begin{figure}
\label{fig_rm}
\psfig{figure=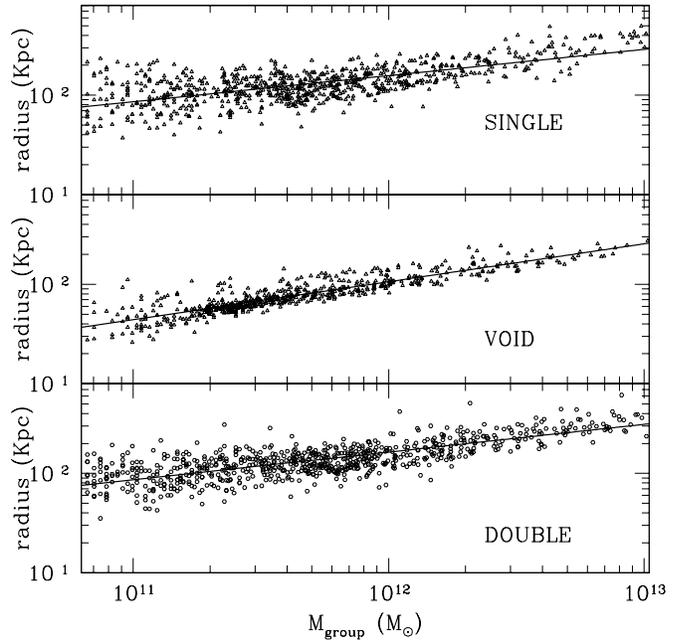,width=9.0cm}
\caption{Relationship betwen truncation radius and mass. Data are fitted using
a power-law relationship: $r_{t}=c M^{\alpha}$. Least-squares fit values
are: $c_{SINGLE}=1.0789\times 10^{-1}, \alpha_{SINGLE}=0.2637\pm 0.1335; 
c_{VOID}=2.7989\times 10^{-3}, \alpha_{VOID}=0.3816\pm 0.0789;
c_{DOUBLE}=7.3451\times 10^{-2}, \alpha_{DOUBLE}=0.2791\pm 0.1193$.
}  \end{figure}

We will then look for solutions in
the interval $2.91 \leq \zeta_t \leq 59.5$. Within these limits the function $\Phi(\zeta
)$ is monotonically decreasing and the solution of eq.~\ref{eq_glob} is certainly
unique. Note that the TIS solution is unstable for $\zeta_t > 34.2$
\citep{1999MNRAS.307..203S, 1962spss.book.....A, 1968MNRAS.138..495L}, so our
choice of the upper limit will allow us to verify {\em a posteriori} this prediction.
Equation \ref{eq_glob} can be inverted w.r.t. \zt\, given the left-hand side,
because for each group all the quantities in the left-hand side are computed by the
group finder. As for the truncation radius \rt\,we adopt the 
groups' radius as computed by \skid, which coincides with the virial radius
$r_{VIR}$.\\
An interesting feature of the TIS solution, which is evident from Figure~8, is
that in order to invert eq.~\ref{eq_glob} to find \zt, the value of the left-hand side
must lie within a rather small range of values: $ 2.75\leq
GM_{0}/r_{t}\sigma_{v}^{2}\leq 4.56$. As we can see from Figure~9 for the case
of the DOUBLE region, this is not the case for all the halos, even for those
halos closely verifying the \svm\, relationship. For instance, out of 827 halos
identified by \skid\, within the DOUBLE region, only 382 lie within the region
for which eq. \ref{eq_glob} can be inverted. For those halos for which
eq.~\ref{eq_glob} can be inverted, we are able to determine \zt\, and to
compare it with the predictions of the minimum-energy TIS model. The results of
this exercise give us some insight into the properties of these halos. In
Figure~6 we plot the relationship between \zt\, and the mass
for halos in the VOID and DOUBLE region (the behaviour of halos in the SINGLE
region is similar to that of those in DOUBLE). The difference between halo
properties in these two regions is striking. In the DOUBLE region there is no
clear relationship between \zt\, and mass, but we do not either find a
clustering around the value $\zeta_{t}=29.4$, characterizing the minimum-energy
TIS solution as suggested by \citet{1999MNRAS.307..203S}. On the other hand,
there seems to be a relationship between \zt\, and mass for halos in the VOID
region, although with a rather large dispersion,  particularly for halos having
$M_{0} \la 3\times 10^{11}h^{-1}M_{\sun}$. \\ 

\begin{figure}
\label{fig_phi}
\psfig{figure=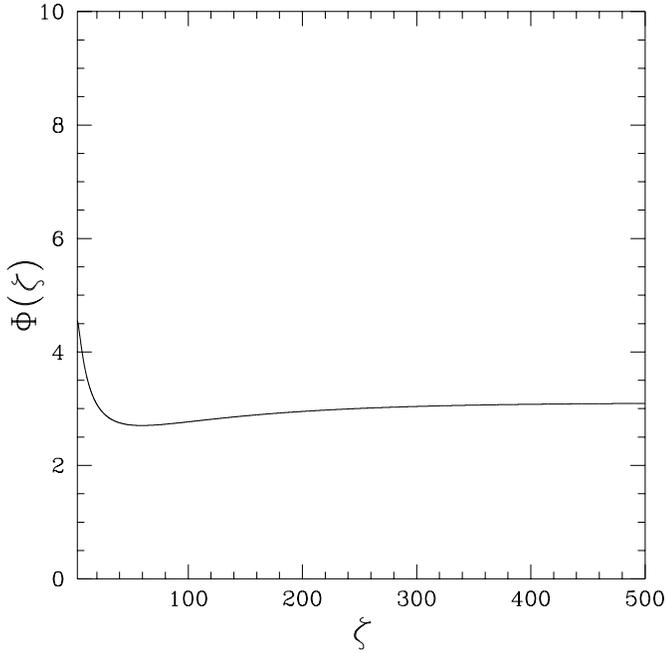,width=9.0cm}
\caption{Plot of $\Phi\zeta_t )$. We start plotting from $\zeta=0.97$, where the
function has an absolute maximum. The vertical scale is the same as that adopted in
Fig.~7.} 
\end{figure}

A very interesting property of halos in Figure~6 is
that we do not find halos having $\zeta_{t}\geq 34.2$, the upper limit for
gravothermal instability for TIS halos, although our upper limit for the \zt\,
values extends up to $\zeta_{t}\leq 59.5$. We do however also find a few halos
having  $\zeta_{t}< 4.738$, the critical value below which the total energy $E
= T + K > 0$ and the TIS solution can not exist \citep{1999MNRAS.307..203S}. It
is important to remember that the halos we find in a simulation are not
spherical and not even ``ideal'', being a discrete realization of some
equilibrium state, so the above quoted bounds can not be taken literally.\\ At
first sight, it may seem curious that only a fraction of all the halos (46\% in
the  DOUBLE and 37\% in the VOID, respectively) have values of \sv\, and
$M_{0}$ for which eq.~\ref{eq_glob} can be solved. The obvious interpretation
is that only a fraction of halos have reached equilibrium, even at the end of
the simulation; but it also remains the possibility that these
``out-of-equilibrium'' halos relax to an equilibrium state different from any
of the three considered in the present work. 

\begin{figure} \label{fig_glob}
\psfig{figure=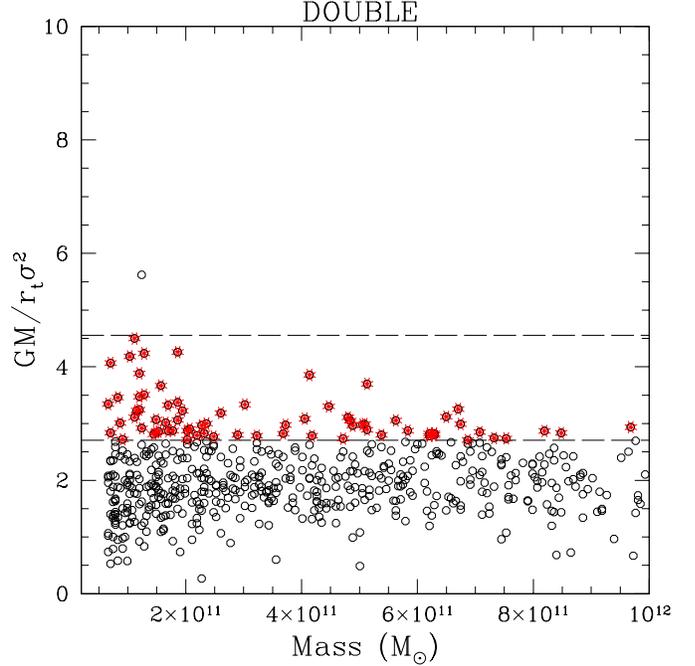,width=9.0cm} \caption{Measured values of
$GM_0/r_{t}\sigma_{v}^{2}$ versus Mass for the region of the DOUBLE cluster.
Groups marked with a star lie within the allowed region where solutions
of eq. \ref{eq_glob} exist, and the corresponding value of \zt can be
found.}  
\end{figure}

Finally, in Figs.~10-11 we plot the \svm\, relation of those
halos for which a \zt\, can be found, for the DOUBLE and VOID regions, respectively. 
Note that
for these halos the intrinsic dispersion is even smaller than in Figs.~3 and
4. These  halos can be then regarded as very near to a TIS equilibrium state.
Note also that the
coefficient of eq.~\ref{eq_sv} for the TIS solution has been computed for the
minimum energy TIS solution. In fact, from eq. (98) of \citet{1999MNRAS.307..203S}
we see that this coefficient would depend on $\zeta$:
\be
\label{eq_cz}
\sigma_{v}^{2}=\frac{\left( 3\pi G\right)}{5}^{2/3}
\frac{\alpha(\zeta )}{\alpha(\zeta )-2} H_{0}^{2/3}(1+z_{coll})M_{0}^{2/3}
\ee
where: $\alpha(\zeta )=3\tilde{M}(\zeta )/\zeta^{3}\tilde{\rho}(\zeta )$. As we can
see from the figures, the scatter induced by this dependence is very small and less
than the intrinsic Poisson scatter.

\begin{figure}
\label{fig_zm_double}
\psfig{figure=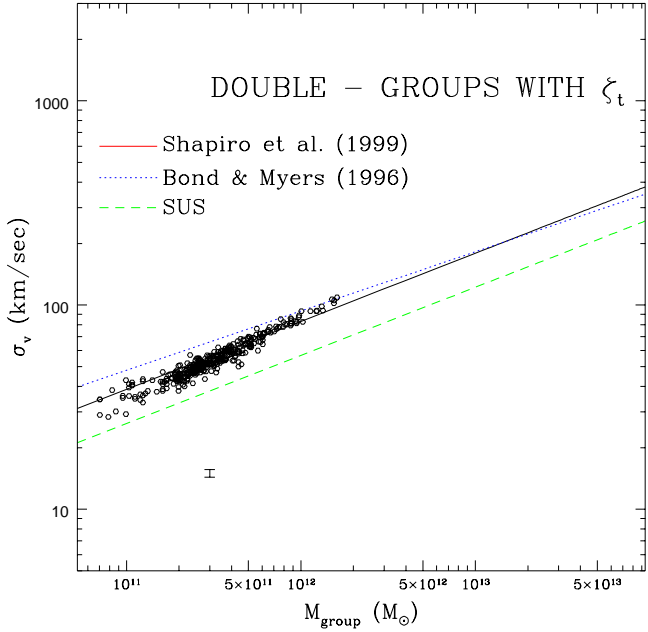,width=9.0cm}
\caption{The \svm\, relation for groups in the DOUBLE region for
which the truncation radius can be computed by inverting eq. \ref{eq_glob}. The
scatter is less than in the analogous plot for all the groups found by \skid\, in the 
DOUBLE region, Fig.~3. The error bar in the lower left part of the plot represents
the uncertainty in the zero point of the best-fit curve induced by the scatter in 
observed values of $\zeta_{t}$. Notice that it is less than the observed scatter. } 
\end{figure}

\begin{figure}
\label{fig_zm_void}
\psfig{figure=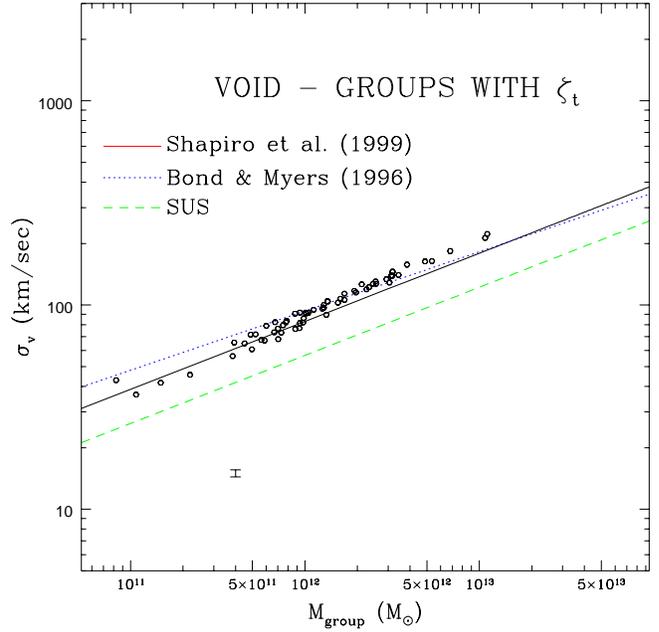,width=9.0cm}
\caption{Same as Fig.~10 for the VOID region. Note that the scale of the axes is the
same as in Figure~5.} 
\end{figure}

\subsection[]{Probability distribution of the spin parameter} \label{pdosp}

The angular momentum distribution is an interesting statitistics, because the
halo angular momentum originates from gravitational interactions between the
collapsing region and its environment. Following \citet{1971A&A....11..377P} and
\citet{1979MNRAS.186..133E} we will present results for the distribution of the
spin parameter $\lambda$ defined as: 
\be \lambda=\frac{L |E|^{1/2}}{GM^{5/2}}  \label{lambda} 
\ee 
where $L$ and $|E|$ are the angular momentum and the total
energy of each halo, respectively. The calculation of $|E|$ is not free of
ambiguities, because in order to compute the potential energy $W$ one should
take into account the fact that the halo is not isolated, i.e. one should also 
account for the contribution from the environmental gravitational  field, and
this is not currently done by any of the group finders we have adopted. For
this reason, we show in Figure~12 the spin probability distribution \pld\,
computed only for TIS halos in the three regions, i.e. for those halos
verifying eq.~\ref{eq_glob}. As we have seen in the preceding paragraphs, these
halos seem to verify the \svm\, relationship with a much smaller scatter than
halos selected by any group finder, so we regard them as our fiducial
equilibrium halos. Combining eqs. 44 and 45 from \citet{1999MNRAS.307..203S} we
find that for a TIS halo the total energy $E$ is connected to the potential $W$
by: 
\be 
E = -\frac{2-\alpha}{2(\alpha - 1)}W \label{eq:tis_1}
\ee
The potential $W$ for these halos is then computed exactly, i.e. by summing the
contribution from each particle in the simulation. Note that the parameter
$\alpha $ depends on the dimensionless truncation radius \zt\, which can be
evaluated only for TIS halos.\\
One immediately notices that \pld\, seems to depend on the environment. It has been shown in recent work that a very good fit to
\pld\, is given by a lognormal distribution \citep{1997ApJ...482..659D,
1998MNRAS.295..319M}:
\be
\lambda{\rm P}(\lambda ) = \frac{1}{\sqrt{2\pi}\sigma_{\lambda}}\exp\left\{
-\frac{\ln^{2}(\lambda/\langle\lambda\rangle}{2\sigma_{\lambda}^{2}}\right\}
d\lambda \label{eq:fit1}
\ee
\citet{1998MNRAS.295..319M} suggest that Eq.~\ref{eq:fit1} with
$\langle\lambda\rangle = 0.05, \sigma_{\lambda}=0.05$ provides a good fit to
the probability distribution of all halos, independently of the environment.
What we observe is, on the other hand, that all 
the three observed distributions seem to be reasonably well fitted by
eq.~\ref{eq:fit1}, but the values of the fitting parameters are
certainly very different from those mentioned above. Moreover, for those
halos selected with \afof, the distributions seem to be all consistent with
each other but not well fitted by the lognormal model (Figure~13).\\
This discrepancy may appear puzzling. However, Figures~12 and 13 cannot be
compared, because of two reasons. First, the total energy $E$ entering the
definition of $\lambda$ was evaluated directly in Figure~12, while it was
estimated from eq.~\ref{eq:tis_1} in Figure~8. The procedure we have adopted
to extract {\em fiducial} TIS halos does actually produce a sample which
verifies the \svm\, relationship with a much less statitistical noise than the
parent sample. But there is an even stronger reason which makes the comparison
doubtful: the total number of halos extracted using \afof\, is much less than
that obtained using \skid . Moreover, the extent over $\lambda $ of the spin
probability distribution is smaller than for \skid\, as is evident from a
comparison of the figures. If we take these differences into account, we do
not see any significant difference among the distributions in the VOID
region. For all these reasons, we can conclude that there is a dependence of
the spin probability distribution \pld\, on the environment {\em only} for TIS
halos. It would be interesting to speculate about the physical mechanisms
producing this dependence, and we hope to be able to address this question in
further work.

 \begin{table}
 \caption{Fits of \pld\, with lognormal distribution, for TIS halos.}
 \label{pfit1}
 \begin{tabular}{lccc}
  Region & Median & $\langle\lambda\rangle$ & $\sigma_{\lambda}$ \\
 & & & \\
  VOID & 0.018 & 0.018 & 0.5 \\
  DOUBLE & 0.018 & 0.03 & 0.9 \\
  SINGLE & 0.051 & 0.07 & 1.4 \\
  All (SKID) & 0.06 & 0.07 & 0.65 \\
 
 \end{tabular}
\end{table}
Before closing this section, we would like to remind the reader that recent 
theoretical calculations predict a rather large distribution in the average
values and shape of \pld\, , with a rather marked dependence on the peak's
overdensity \citep{1996MNRAS.282..436C} or on the details of the merging
histories \citep{1998MNRAS.301..849N,2001astro-ph...0105349}. A direct comparison of our results with
the conditional probability distribution ${\rm P}(\lambda | \nu)$ of
\citet{1996MNRAS.282..436C} is made difficult by the fact that the relationship
between the linear overdensity $\nu$ and (for instance) the final mass of the
halo turns out to be quite noisy \citep[Fig. 10]{2000MNRAS.311..762S}, so it is
not possible to ``label'' unambigously each halo with its initial overdensity.
However, one could hope to further increase the number of halos by further
diminishing the softening length, and we hope to get a better statistics from
future simulations which would help us to address also the latter points.

\begin{figure}
\label{fig_lambda_dist}
\psfig{figure=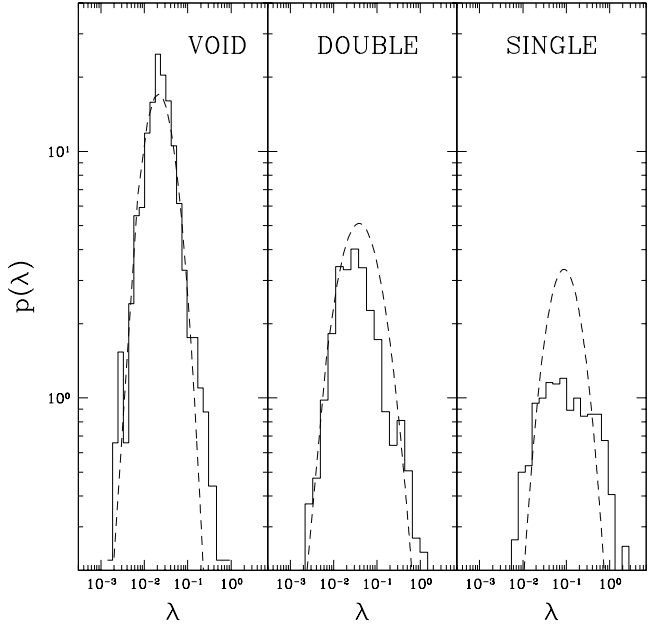,width=9.0cm}
\caption{Probability distribution of the spin parameter for the three
regions, for TIS halos. The dashed curves are the best-fit approximations
obtained using the lognormal distribution adopted by
\citet{1998MNRAS.295..319M} (eq. 15). Histograms and fitting  curves are
normalized to unity.}  \end{figure}

\begin{figure}
\label{fig_lambda_dist_afof}
\psfig{figure=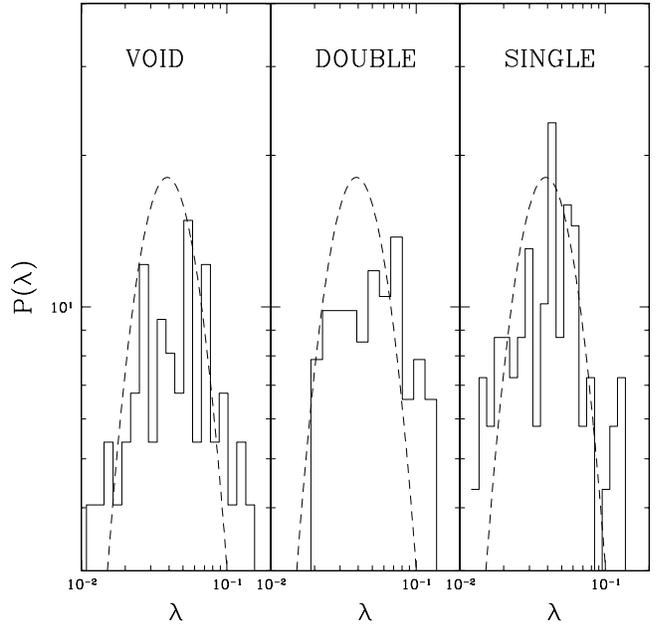,width=9.0cm}
\caption{Same as Figure~14 but for halos selected using \afof . The dashed
curve is a lognormal distribution with $\rangle\lambda\langle=0.5,
\sigma_{v}=0.05$.}
\end{figure}

\subsection[]{Density profiles of massive halos} \label{dp}
As we already mentioned in the Introduction, even the most massive halos we find in
this simulation using \skid\, do not contain enough particles to allow a reliable
determination of the density profile. This is clearly visible from Figure~13,
where we plot the profiles of the four most massive halos extracted from the
DOUBLE cluster region. None of these halos lies in the integrability strip of
Figure~11, so the best fit TIS profiles displayed as continous curves have been
obtained by least square fittings, where we have varied $\rho_{0}$ and $r_{0}$.

\begin{figure}
\label{rho_vs_r_4w}
\psfig{figure=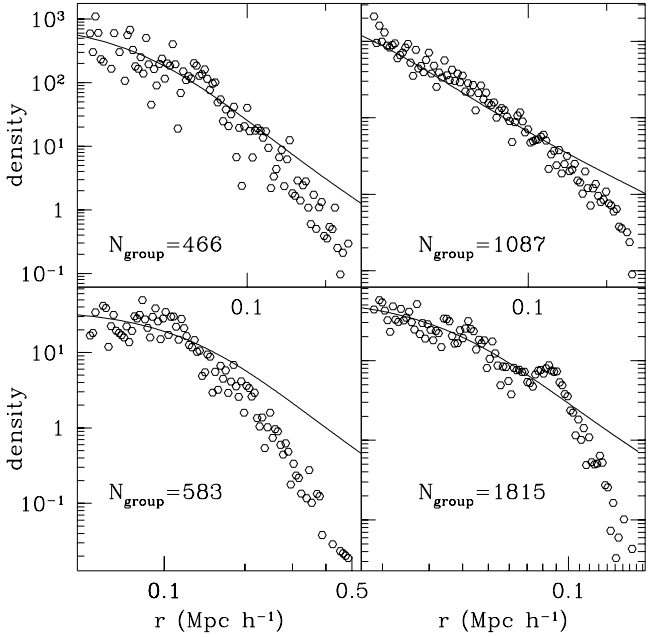,width=9.0cm}
\caption{Density profiles for the four most massive halos within the DOUBLE
region. The number of particles contained in each halo is shown. Best fit solutions
using the minimum energy TIS solution by \citet{1999MNRAS.307..203S} are shown.} 
\end{figure}
The distinguishing feature of the TIS density profile, when compared with the
universal density profile of \citet{1996ApJ...462..563N, 1997ApJ...490..493N} is the
presence of a central core. Although the minimum energy TIS profiles fit reasonably
well the central regions, they fall out too gently at distances larger than a few times
the core radius, and in no case we can find a reasonably good overall agreement. It
would be hazardous to draw any conclusion from this comparison, in view of the
above mentioned poor resolution. However, a reasonable explanation for the sharp decline of 
the density profiles is tidal stripping, which should be effective at a few times
the core radius.

\section[]{Conclusion} \label{conc}
The properties of galaxy-sized halos we have considered in this paper seem to be
very constraining for halo collapse and equilibrium models. However, none of the
equilibrium models considered (neither the {\em minimum energy TIS} model)
seems to be able to give a comprehensive description of our findings.
We would like to summarize now our findings and to point to some
controversial issues that they pose.\\
First of all, the \svm\, statistics seems to be a sensitive tool to discriminate among
different halo equilibrium models. This statistics is easy to evaluate, because it
relies on global quantities, and it can then be applied to samples of halos. In this context, 
it is more difficult to discriminate models using statistics like the density
profile, which would require a considerably larger mass range in order to give
reliable results \citep[see for instance][]{2000ApJ...529L..69J}.\\
Models for the \svm \, statistics based on the NFW density profile seem to
be only marginally consistent with simulation data. The r\^{o}le of the
anisotropy parameter in this context does not seem to be crucial: it is the
slope of the radius-mass relationship for these halos which seems to mostly
affect the normalisation of the \svm \, .\\
\noindent
 As we have seen, the TIS model seems to offer a very good
quantitative framework to explain the \svm\, statistics, even in the VOID
region where the slope of the relationship is very different from that
predicted by the minimum energy TIS model of \citeauthor{1999MNRAS.307..203S}
The fact that a model based on the hypothesis that halos have a {\em finite
extent} provides a good description should not come as a surprise. Halos
forming in clusters experience a complex tidal field  originating from
neighbouring halos and from the large scale web in which they are embedded. The
tidal radii of the environments within which they lie, although often larger
than the mean distance, could limit the extent of halos. A theoretical
treatment of the growth of the angular momentum is complicated by the fact that
the distribution of the torques induced by nearby halos depends on clustering
\citep{1992ApJ...392..403A}. However, we believe that it would be difficult to
think that the truncation is a numerical artifact due to the finite mass
resolution: were this the case, we should expect the same relationship between
truncation radius $r_{t}$ and mass in all the three regions, but this is
clearly not the case.\\ We have already noted the fact that the truncation
radii we find are always less than the critical value for the onset of
gravothermal instability, $\zeta_{crit}=34.2$. This leads us to think that this
instability is at work in our simulations, but in order to investigate this
issue one would need simulations with a dynamical range at least 3 orders of
magnitudes larger than those used in this simulation.\\ 
Concerning the dependence of the distribution of spin parameters on the
environment, we find that halos selected using \afof\, do not show any
dependence on the environment (the same holds for halos selected using
\skid\,), but if we select subsamples of {\em fiducial} TIS halos, we do find a
dependence of the properties of \pld\, on the environment. In particular, this
fact seems to be at odds with the recent investigation by
\citet{1999MNRAS.305..357S}, who find that the observed distribution of the
spin parameter for a large homogeneous sample of spirals is well described by a
lognormal distribution with $\langle\lambda\rangle\approx 0.05$ and a variance
$\sigma_{\lambda}\approx 0.36$. This result is in contrast also with other work
mentioned above \citep{1992ApJ...399..405W, 1995ApJ...439..520E}. If confirmed
by further investigations, this discrepancy could suggest that there is
probably some systematic trend in the way the angular  momentum of the luminous
discs is connected with that of the halo, which is not accounted for by the
models of \citet{1999MNRAS.305..357S}.\\ Last but not least, it is important to
stress that \citet{1999MNRAS.302..111L} conclude that ``Only the mass
distribution varies as a function of environment. This variation is well
described by a simple analytic formula based on the conditional
Press--Schechter theory.  We find no significant dependence of any other halo
property on environment...''. In comparing their results with ours, we must
keep in mind that we have followed a very different procedure from theirs,
because we have {\em prepared} a simulation using constrained initial
conditions with the purpose of obtaining a final configuration containing
certain features (i.e. a double cluster and a void). Although our simulation
box is not an ``average'' region of the Universe, it is certainly a
representative one. We stress again the fact that all the halos from underdense
regions in our simulation come from a void, and not from the outer parts of
clusters. \citeauthor{1999MNRAS.302..111L}, on the other hand, seem to take
their halos from all the volume and group them according to the overdensity of
their parent regions. We think then that a direct comparison between the
results of these two different investigations would be misleading, given the
complementarity of our approaches.

\section[]{Appendix}
We give a full derivation of equation~\ref{eq_glob}. The starting
point is the virial theorem for systems with boundary pressure terms, as given in
eq. 41 from \citet{1999MNRAS.307..203S}:
\be
0=2K+W+S_p \label{eq_vt}
\ee
In the above equation the kinetic energy $K$ can be rewritten in terms of the 1-D
velocity dispersion:
\be
K = \frac{M_{0}\langle v\rangle^{2}}{2}=\frac{3}{2}M_{0}\sigma_{v}^{2} \label{eq_k}
\ee
The potential energy term $W$:
\be
W = 4\pi G\int_{0}^{\infty} \rho M(r)rdr\equiv 4\pi G\int_{0}^{r_{t}} \rho
M(r)rdr
\ee
can be rewritten in terms of global quantities and of the dimensionless
radius \zt:
\be
W =-\frac{GM_{0}^{2}}{r_{t}}\frac{\zeta_{t}\Psi(\zeta_{t})}{M^{2}(\zeta_{t})}
\label{eq_w}
\ee
where we have defined:
\[
\Psi(\zeta_{t}) = \int_{0}^{\zeta_{t}}d\zeta\zeta\tilde{\rho}(\zeta)\tilde{M}(\zeta)
\]
Finally, $S_p$ is a surface term which arises from the constraint that the system has a
finite radius, and is given by (Shapiro et al., eq. 43):
\be
S_p = -4\pi r_{0}^{3}p_{t}\zeta_{t}
\ee 
We are adopting here the same notation as Shapiro et al., so that $r_{0}$ and $p_{t}$ are
the core radius and an external ``pressure'' term,  respectively. Using eqs. 34
and 38 from Shapiro et al., the latter equation can be rewritten in terms of the dimensionless 
integrated mass and density:
\be
S_p =
-\frac{M_{0}}{\tilde{M}_{t}(\zeta_{t})}\tilde{\rho}(\zeta_{t})\sigma_{v}^{2}
\label{eq_sp}
\ee
Substituting eqs.~\ref{eq_k},~\ref{eq_w},~\ref{eq_sp} into eq.~\ref{eq_vt} we get:
\be
3M_{0}\sigma_{v}^{2}
 = -\frac{GM_{0}^{2}}{r_{t}}\frac{\zeta_{t}\Psi(\zeta_{t})}{M^{2}(\zeta_{t})} -
\frac{M_{0}}{\tilde{M}_{t}(\zeta_{t})}\tilde{\rho}(\zeta_{t})\sigma_{v}^{2}
\ee
from which we get the desired equation.
\section*{Acknowledgments}

V.A.-D. is grateful to prof. Paul Shapiro and P. Salucci for useful comments.
Edmund Berstchinger and Rien van de Weigaert are
gratefully acknowledged for providing their constrained random field code.

\bibliographystyle{/home/antonucc/t3e/16ml/paper1/tex/mnras/v1/apj}
\bibliography{/home/antonucc/t3e/16ml/paper1/tex/mnras/v1/paper16ml_1}

\bsp

\label{lastpage}

\end{document}